\documentclass[preprint,showkeys,aps,prb]{revtex4}
\usepackage{amsmath}
\usepackage[dvips]{graphicx}
\usepackage{setspace}

\begin{document}

\title{
Nonequilibrium Molecular Dynamics, Fractal Phase-Space \\
Distributions, the Cantor Set, and Puzzles Involving \\
Information Dimensions for Two Compressible Baker Maps. \\
}

\author{
William Graham Hoover and Carol Griswold Hoover \\
Ruby Valley Research Institute                  \\
Highway Contract 60, Box 601                    \\
Ruby Valley, NV 8983                            \\
}

\date{\today}

\keywords{Chaos, Lyapunov Exponents, Irreversibility, Random Walks, Maps, Information Dimension}

\begin{abstract}
Deterministic and time-reversible nonequilibrium molecular dynamics simulations typically generate
``fractal'' [ fractional-dimensional ] phase-space distributions. Because these distributions and
their time-reversed twins have zero phase volume, stable attractors ``forward in time'' and unstable
(unobservable) repellors when reversed, these simulations are consistent with the Second Law of
Thermodynamics. These same reversibility and stability properties can also be found in compressible
Baker Maps, or in their equivalent random walks, motivating their careful study.  We illustrate these
ideas with three examples: a Cantor-Set Map and two linear compressible Baker Maps, N2$(q,p)$ and
N3$(q,p)$. The two Baker Maps' Information dimensions estimated from sequential mappings agree while
those from pointwise iteration do not, with the estimates dependent upon details of the approach to
the maps' nonequilibrium steady states.

\end{abstract}

\maketitle

\section{Nonequilibrium Molecular Dynamics Generates Fractals}

The computers developed for the National Laboratories were first applied to manybody
problems in the 1950s.  At Los Alamos in 1953, Fermi, Pasta, and Ulam\cite{b1} described
the incomplete equilibration of one-dimensional waves in anharmonic chains.
Soon afterward, at Livermore, Berni Alder and Tom Wainwright simulated the motion
of systems of several hundred hard disks and spheres\cite{b2}. At Brookhaven George
Vineyard and his coworkers studied ``realistic'' atomistic models of the impact
of high-energy radiation on models of simple metals shortly thereafter\cite{b3}.  All
of these atomistic simulations were developed based on classical Newtonian mechanics with
short-ranged pairwise-additive forces.  ``Large'' simulations involved several
hundred discrete particles. A generation later simulations with millions of particles were
possible. {\bf Figure 1} shows a typical simulation from our 1989-1990 visit to Japan. These
indentations of amorphous Stillinger-Weber silicon, using two different indentor models,
generate plastic flow near the indentors\cite{b4}.  Trillion-atom simulations are
feasible in 2020.

In 1984 Shuichi Nos\'e had announced a revolutionary method for imposing specified temperatures
and pressures on molecular dynamics simulations\cite{b5,b6}. His modification of Hamiltonian
mechanics was designed to replicate Gibbs' isothermal and isobaric ensembles. Equilibrium
distributions had been formulated by Gibbs' statistical mechanics prior to the close of the
19th century. To match Gibbs' results Nos\'e found it necessary to introduce a ``scaled''
time which had the drawback of introducing wild fluctuations in the dynamics. Hoover helped
develop these ideas into practical numerical algorithms\cite{b7} which avoided time-scaling.
The simplest problem to which the Nos\'e-Hoover approach can be applied is the harmonic
oscillator with unit mass, force constant, and temperature :
$$
\dot q = p \ ; \ \dot p = -q -\zeta p \ ; \ \dot \zeta = p^2 - 1 \ [ \ {\rm NH} \ ] \ .
$$
Here $(q,p)$ are the oscillator coordinate and momentum.  The time-averaged kinetic
temperature, $\langle \ p^2 \ \rangle$ is controlled (``thermostatted'') by the
time-reversible friction coefficient $\zeta$. The reader can verify, as was pointed
out in Reference 7, that applying the steady-state continuity equation for the flow
in $(q,p,\zeta)$ space gives Gibbs' canonical distribution for the oscillator together
with a Gaussian distribution for the friction coefficient $\zeta$ :
$$
(\partial f/\partial t) = 0 = -\partial (f\dot q)/\partial q -\partial (f\dot p)/\partial p
-\partial (f\dot \zeta)/\partial \zeta \ \longleftrightarrow \ 
f(q,p,\zeta) \propto e^{-(q^2 + p^2 + \zeta^2)/2} \ .
$$
This idea and its isobaric analog have become standard approaches to equilibrium molecular
dynamics simulations for a wide variety of systems both large and small.
 
Although pairwise-additive potentials might seem an oversimplification, work
earning Nobel Prizes in chemistry (1986 and 2013), carried out by Dudley
Herschbach and Martin Karplus and their colleagues, showed otherwise\cite{b8}.
Straightforward classical solutions of pairwise-additive motion turned out to
be quite useful in interpreting and predicting the properties of molecules both
simple (hydrogen and various alkalai halides) and complex (proteins).  In most
isoenergetic dynamics simulations atomistic trajectories are generated using the
``Verlet'' or ``St\"ormer'' algorithm, with its roots going all the way back to
Newton.  This algorithm expresses the ``next'' coordinate value, $x_{t+dt}$ in
terms of the previous and current coordinates along with the current value of
the acceleration $\ddot x_t$ :
$$
x_{t+dt} = 2x_t - x_{t-dt} \equiv dt^2(F_x/m)_t \longleftarrow \ddot x = (F_x/m) \ .
$$
Similar time-reversible algorithms have been developed for isothermal and
isobaric systems\cite{b9}.

Though not time-reversible, fourth-order Runge-Kutta integration of the system of
first-order motion equations,
$$
\{ \ \dot x = (p_x/m) \ ; \ \dot p_x = F_x \ \} \ ,
$$
provides better accuracy at a fixed timestep $dt$, particularly for the velocities.
Typical equilibrium simulations, based on Newtonian or Hamiltonian or Nos\'e-Hoover
mechanics, use periodic boundary conditions. Series of such simulations can be used
to generate ``equations of state'', temperature and pressure as functions of energy
and density.

Nonequilibrium simulations such as Vineyard's radiation-damage studies, or the
simulation of planar shockwaves, require the implementation of special boundary
conditions capable of imposing velocity and temperature differences across systems
of interest. An early discovery motivating quantitative atomistic simulations was
the finding that the width of strong shockwaves is on the order of the size of
molecules so that details can be modelled reasonably well with only a few thousand
particles. The atomistic and continuum descriptions of strong shockwaves were in
rough, ten percent, agreement with one another\cite{b10}.

By 1987 a significant difference between equilibrium and nonequilibrium steady
states had come to light. Simple nonequilibrium simulations were shown to produce
fractal ( fractional-dimensional ) phase-space distributions, with a negligible
phase-space volume relative to corresponding higher-dimensional equilibrium Gibbs'
distributions, such as the microcanonical and canonical ensembles\cite{b11,b12,b13}.
About the simplest steady-state mechanical problem results when heat is driven through
a Nos\'e-Hoover harmonic oscillator exposed to a temperature gradient\cite{b14,b15}.
Where the maximum value of the temperature gradient is $\epsilon$ the three motion
equations (for the coordinate $q$, the momentum $p$, and the current-driving friction 
coefficient $\zeta$), are as follows :
$$
\dot q = p \ ; \ \dot p = - q - \zeta p \ ; \ \dot \zeta = p^2 - T(q) \ ; \
T(q) = 1 + \epsilon \tanh(q) \ [ \ {\rm NH} \ ] \ .
$$

Here again, the oscillator mass, force constant, and mean temperature have all been
chosen equal to unity, but the temperature gradient can generate nonequilibrium steady
states. The three coupled equations give rise to a wide variety of solutions. Three
such solutions, all for a maximum temperature gradient $\epsilon = 0.42$, are shown in
{\bf Figure 2}. There we see a dissipative limit cycle as well as two conservative
tori.  Chaotic solutions are also accessible to the model.  More complicated
mechanical models, with two friction coefficients, $\zeta$ and $\xi$, and the same
coordinate dependence of the temperature, $T(q) = 1 + \epsilon \tanh(q)$, can generate
{\it ergodic} fractal distributions.  In ergodic systems the same longtime steady-state
solutions obtained apply for any initial condition (``almost any'' for the mathematically
minded). Two examples are shown in {\bf Figure 3}.  The Hoover-Holian\cite{b16} and 
Martyna-Klein-Tuckerman\cite{b17} motion equations which generated them are as follows :
$$
\dot q = p \ ; \ \dot p = - q - \zeta p - \xi p^3 \ ; \ \dot \zeta = p^2 - T(q) \ ; \
\dot \xi = p^4 - 3p^2T(q) \ ; \ [ \ {\rm HH} \ ] \ ;
$$
$$
\dot q = p \ ; \ \dot p = - q - \zeta p \  ; \ \dot \zeta = p^2 - T(q) - \xi \zeta \ ; \
\dot \xi = \zeta^2 - T(q) \ ; \ [ \ {\rm MKT} \ ] \ .
$$
At equilibrium, with $T \equiv 1$, the two Hoover-Holian thermostat variables control both
the second and the fourth moments of momentum. The Martyna-Klein-Tuckerman $\xi$ controls
the distribution of the other thermostat variable $\zeta$. Detailed investigations of these
oscillator problems at equilibrium show that both the HH and the MKT dynamics give the same
ergodic distribution, including Gibbs' canonical Gaussians in both $q$ and $p$.
$$
f(q,p,\zeta,\xi) \propto e^{-(q^2+p^2+\zeta^2+\xi^2)/2} \ .
$$

The distributions away from equilibrium can be qualitatively different! The fractal ergodic
nature of nonequilibrium states generated with a variable $T(q)$ provides a simple explanation
for the irreversible behavior (despite time-reversible motion equations) described by the
Second Law of Thermodynamics.  When the nonequilibrium motion equations are solved, the fractal
strange attractor states which result are not only vanishingly rare. In addition their
time-reversed states (with the momenta and the friction coefficients' signs changed) describe
mechanically-unstable fractal repellors. Because such repellors correspond to {\it exploding}
rather than {\it collapsing} phase-space objects they are {\it unobservable}, having negative
transport coefficients and entailing entropy destruction rather than production!\cite{b11,b12,b13}.

From a thermodynamic standpoint phase-volume increase corresponds to heating and decrease to cooling.
Because irreversible processes necessarily increase heat it follows that a stationary nonequilibrium
distribution must extract net heat, leading to the formation of a fractal attractor.  A reversed
process converting heat to work, a repellor, is outlawed by both the Second Law of Thermodynamics and
by computational instability in time-reversed numerical simulations\cite{b12}.

We consider fractal distributions in more detail in the following Sections.  We have already seen
that in the 1980s nonequilibrium molecular dynamics led to the characterization of fractal
(fractional dimensional) distributions. These are qualitatively different to Gibbs' smooth equilibrium
distributions.  Because the mathematics of fractals and their geometric characterization is
interesting and sometimes paradoxical we highlight stimulating research areas well-suited to
student exploration in what follows. In Section II we begin with the simplest fractal, the Cantor
set, and a description of its fractal dimensionality.  Section III takes up time-reversible
compressible Baker Maps, where phase-volume changes model nonequilibrium heat transfer.  The one-way
nature of these maps (``Time's Arrow'') is a direct geometric analog of the Second Law conversion of
work to heat.  Finally we consider Kaplan and Yorke's relation linking the fractal information
dimension to the Lyapunov exponents. We show that their conjectured equality between the information
dimension and the Kaplan-Yorke dimension, $D_{KY} \stackrel{?}{=} D_I$, is precisely true for one
Baker map, N3, and apparently false for its very similar twin, N2. This surprise was completely
unexpected.  It richly deserves further study.

\section{The Cantor Set and the Fractal Information Dimension}

The simplest fractal is arguably the ``Middle-Third'' ``Cantor Set''. The Middle-Third description
suggests one of the several means for constructing the set: Begin with the unit interval
$[0 \ {\rm to} \ 1]$; Discard the middle third $[(1/3 \ {\rm to} \ (2/3)]$ leaving two intervals,
$[(0/3) \ {\rm to} \ (1/3)]$ and $[(2/3) \ {\rm to} \ (3/3)]$; Discard the middle third of those
two, leaving four intervals of length (1/9); Finally, imagine the limiting set of points remaining
after an infinite number ($\aleph_0$, the number of integers) of removal stages.

A more elegant alternative description of this same Cantor Set, or ``Cantor Dust'', is the set of
numbers on the unit interval whose ternary representation is composed wholly of 0s and 2s.  An
example set member is the base-3 number 0.20220000... = 2/3 + 0/9 + 2/27 + 2/81 + 0 = 62/81. Because
each of $\aleph_0$ digits of the Cantor set can be either a 0 or a 2 the (likewise infinite) number
of Cantor-Set members is $2^{\aleph_0} \equiv \aleph_1$. The fact that the continuum itself, when
expressed in binary base-2 rather than ternary base-3, has likewise $\aleph_1$ members, all the binary
combinations of 0s and 1s, seems paradoxical.  The continuum has no holes while the well-named Cantor
Dust has nothing but! Does it really make ``sense'' to accept the notion that the members of the continuum
and the Cantor Set are equinumerous?  Worse still -- how sound is the notion that the number of members
of the continuum is invariably $\aleph_1$, independent of the continuum's dimensionality, one, two, three,
\dots ? These troubling counter-intuitive aspects of Cantor's ideas (and the undecidability of the
Continuum Hypothesis) suggest an aesthetic-but-inapplicable branch of mathematics.  Nevertheless let us
pursue a descriptive approach to dimensionality differences among the various infinite fractal subsets
of continua.

Alfr\'ed R\'enyi described recipes for various fractal dimensions long ago, the fractal dimension,
information dimension, correlation dimension. Expressed in terms of the probabilities $\{ \ p \ \}$
of occupying a set of bins, all of the same size $\delta$, with $\delta$ sufficiently small, the
forms of these three are as follows:
$$
\begin{textstyle}
D_0 = \ln (\sum [p^0])/\ln (1/\delta) \ ; \ D_1 = \sum [p\ln p]/\ln (\delta) \ ;
\ D_2 = \ln(\sum [p^2])/\ln (\delta) \ .
\end{textstyle}
$$
The sums include all occupied bins. The probabilities are normalized, $\sum[p]\equiv 1$, and are
typically proportional to the number of counts or the fractions of the total time associated with
residence in  each of the occupied bins. The fractal (or capacity), information, and correlation
dimension correspond to $D_0$, $D_1$, and $D_2$. The reader can verify that in the case of the
Cantor set these three dimensions are all the same, $D = \ln(2)/\ln(3) = 0.630930$. Notice that
reducing the bin size by a factor of three results in just twice as many occupied bins. Likewise,
coarsening the bin size by a factor of three results in just half as many occupied bins. $D$ plays
the role of a (fractional) dimension: $3^D = 2 \longleftrightarrow D = \ln(2)/\ln(3)$ .

This simplest of fractals sets the stage for studying two interrelated families of nonequilibrium
fractals even simpler than those generated by the conducting harmonic oscillator problems.  The
two families of simpler models are [1] stochastic random walks (usually on the unit interval from
0 to 1) and [2] deterministic time-reversible compressible maps (where we use a rotated $2\times 2$
diamond-shaped domain in order to model time-reversibility and to enhance ergodicity). The expected
equivalence of these models is itself interesting.  The fact that such simple models can lead to
results that are contradictory or paradoxical, despite the long history of their study, is currently
in need of further pedagogical explanation. The Ian Snook Prize for 2020\cite{b18}, to be awarded to
the author(s) best addressing this need, is designed to shed light on these families of fractal problems.

The information dimension $D_I=D_1$, a close relative of Gibbs' statistical entropy, is arguably the most
useful descriptor of fractal point sets or distributions.  Using the $p\ln p$ formula for $\delta = 1/27$
there are eight three-digit members of the Cantor set :
$$
\{ \ 0.000, \ 0.002, \ 0.020, \ 0.022, \ 0.200, \ 0.202, \ 0.220,\  0.222 \ \} \ .
$$
The resulting dimensionality is $\ln(1/8)/\ln(1/27) = \ln(2)/\ln(3) = 0.630930$ .  For any fixed  number
of digits the same distribution-based result is obtained.  A numerical representation of the Cantor set
as an arbitrarily-large set of points can be generated by choosing an initial ``seed'' in the set, like
{\tt C = 2/9} or {\tt C = 62/81}, followed by iteration of the following loop of {\tt FORTRAN} pseudocode:

\pagebreak

\begin{verbatim}
call random_number(R)
if(R.lt.0.5) Cnew = (C/3)
if(R.ge.0.5) Cnew = (C/3) + (2/3)
C = Cnew
write(33,*) it,C
[ Unit Square Generation of the Cantor Set in (R,C) Cartesian Coordinates ]
\end{verbatim}

Here the {\tt FORTRAN random}\_{\tt number} subroutine generates series of random numbers $\{ \ {\tt R} \ \}$ uniformly
distributed between zero and one.  Notice particularly that exactly the same pseudocode describes a random walk with
variable length steps.  Half the time the walker moves left from his present position {\tt C} to {\tt C/3} corresponding
to adding a ternary 0 after the ``decimal'' point. Otherwise, and also half the time, the walker moves to the
right, corresponding to adding in a ternary 2 after the point.  The overall single-step operation shifts the ternary
representation of {\tt C} one digit to the right and then chooses randomly either 0 or 2 to precede it.

Alternatively, a deterministic two-dimensional compressible map can be constructed to generate the Cantor Set in
a rotated $(q,p)$ space with the constant {\tt d} equal to $\sqrt{2}/6$:
\begin{verbatim}
if(q.lt.p) qnew = + (7*q)/6 - (5*p/6) + 5*d
if(q.lt.p) pnew = - (5*q)/6 + (7*p/6) - 1*d
if(q.ge.p) qnew = + (7*q)/6 - (5*p/6) - 5*d
if(q.ge.p) pnew = - (5*q)/6 + (7*p/6) + 1*d
q = qnew
p = pnew
[ Diamond-Shaped Generation of the Cantor Set in (q,p) Space ]
\end{verbatim}
100,000 points generated with these stochastic and deterministic  mappings are illustrated in Figure 4.  Here and
in that Figure, just for convenience in the programming, the $2\times 2$ diamond-shaped domain has extreme values
of $q$ and $p$ of $\pm\sqrt{2}$. In the next Section we elaborate on our preference for the rotated map of
the two-dimensional $(q,p)$ domains rather than the conventional $(x,y)$ unit square.

\section{Information Dimensions for Compressible Baker Maps}
The Baker Map considered by Eberhard Hopf in 1937 provides a simple deterministic model for the dissipative chaos
causing irreversible behavior in the solutions of time-reversible motion equations. By 1987 several examples of
thermostatted molecular dynamics led to the representation of nonequilibrium steady states as fractal structures in
$(q,p)$ (coordinate, momentum) phase space\cite{b11,b12,b13}. A two-panel Baker Map N2, incorporating twofold
changes in the area $dqdp$ is the prototypical example, displayed in {\bf Figures 5 and 6}. This
mapping\cite{b19,b20,b21,b22} follows the equations
\begin{verbatim}
if(q-p.le.-sqrt(2/9)) qnew = + (11/ 6)*q - ( 7/ 6)*p + 14*d
if(q-p.le.-sqrt(2/9)) pnew = - ( 7/ 6)*q + (11/ 6)*p - 10*d
if(q-p.gt.-sqrt(2/9)) qnew = + (11/12)*q - ( 7/12)*p -  7*d
if(q-p.gt.-sqrt(2/9)) pnew = - ( 7/12)*q + (11/12)*p -  1*d
[ Nonequilibrium Two-Panel Baker Map N2 ]
\end{verbatim}
where the constant $d$ is $\sqrt{1/72}$.  The map is irrotational, with unstable fixed points at the top and
bottom of the diamond-shaped domain where $q$ is horizontal and $p$ vertical.  The diamond-shaped domain used
here, $-\sqrt{2} < q,p < +\sqrt{2}$, is purposefully rotated 45 degrees from the usual Cartesian Baker Map. So
as to emphasize the time-reversibility of our compressible Baker Map N2 we choose the diamond-shaped $2\times 2$
domain. Reversing the time corresponds to reversing the sign of the (vertical) ``momentum'' $p$ while keeping
the horizontal ``coordinate'' $q$ unchanged.

This Nonequilibrium Baker Map N2 is ``time-reversible'' in the sense that the inverse mapping, N2$^{-1}$ is given by the
product mapping T{\tt *}N2{\tt *}T. The time-reversal mapping T simply changes the sign of the momentum,
T$(q,\pm p)=(q,\mp p) =$ T$^{-1}(q,\pm p)$. A typical long-time cumulative solution of the N2 mapping is far from homogeneous
but is nonetheless ergodic, with nonvanishing density everywhere within its diamond-shaped domain.  This nonequilibrium
(area-changing) mapping produces no ``holes'' so that its capacity or box-counting or fractal dimension is 2. See the
100,000-point samples of
the mapping and its inverse in {\bf Figure 5}. The N2 mapping is compressive parallel to the line $q=p$ and expansive in
the perpendicular direction. Numerical work indicates that the resulting distribution of points is random in $\tilde{x}$
and remains fractal in $\tilde{y}$ where the orthogonal coordinates $(\tilde{x},\tilde{y})$ occupy a $2\times 2$ square
centered on the origin :
$$
-1 < \tilde{x} = (q-p)/\sqrt{2} \ ; \ \tilde{y} = (q+p)/\sqrt{2} < +1 \ .
$$

For convenience in the measurement of the fractal information and correlation  dimensions and the construction of random
walks in $y$ corresponding to the fractal direction parallel to $q=p$, it is convenient to map the $2\times 2$
$(\tilde{x},\tilde{y})$ square onto the unit $(x,y)$ square :
$$
x \equiv (\tilde{x}+1)/2 \ ; \ y \equiv (\tilde{y}+1)/2 \ .
$$
Then an equivalent set of $(x,y)$ values can be generated by a random walk based on random numbers from the unit interval,
$0 < \{ \ {\tt r} \ \} < +1$
as follows :
\begin{verbatim}
call random_number(r)
x = r
call random_number(r)
if(r.lt.(2/3)) ynew = (0+1*y)/3
if(r.ge.(2/3)) ynew = (1+2*y)/3
y = ynew
[ Two-Panel Nonequilibrium Random Walk in the Unit Square, N2 ]
\end{verbatim}

Because the N2 mapping and this random walk generate the {\it same} long-time distributions of the compressive $y$ variable
the various fractal dimensions\cite{b23,b24,b25} $\{ \ D \ \}$ (box-counting, information, Kaplan-Yorke, and correlation, ... )
are simply related, $D_{\rm map}(q,p) = D_{\rm walk}(y) + 1$.

Careful investigations\cite{b21,b22} of the local densities of points in two-dimensional bins of area $\delta^2 = (1/3)^{2M}$,
with the integer $M$ up to 20, suggested a pointwise fractal information dimension of $1.741_5$. But mapping {\it regions} rather
than {\it points}, and starting with a uniform distribution in the diamond-shaped domain gave a totally different result ! The
information dimension calculated for the same N2 map according to the mapping of regions (areas) rather than by propagation of
a single mapped point can be calculated analytically\cite{b24}.  The result is $D^{\rm N2}_{regions} = 1.789690$ rather than
$D^{\rm N2}_{points} = 1.741_5$ .

\subsection{The Kaplan-Yorke Dimension from Lyapunov Instability}
On the other hand -- forty years ago -- Kaplan and Yorke\cite{b26} conjectured that the fractal information dimensions of
solutions of typical two-dimensional maps are simply related to the solutions' Lyapunov exponents.  Because two-thirds of
the N2-mapped area undergoes a 1.5-fold stretching while one-third undergoes three-fold stretching the larger Lyapunov
exponent is $\lambda_1 = (2/3)\ln(3/2) + (1/3)\ln(3) = +0.636514$. The smaller (negative) Lyapunov exponent describes the
shrinking: $\lambda_2 = (2/3)\ln(1/3) + (1/3)\ln(2/3) = -0.867563$. Kaplan and Yorke reasoned that the information dimension
for such a map is given by
$$
D_I= 1 - (\lambda_1/\lambda_2) = 1.733680 \ ,
$$
a bit less than the estimate from bin-density data\cite{b21,b22} and far from the analytic area-mapping result, 1.789690.

In our efforts to understand these differences we came upon a related N3 mapping, compared to N2 in {\bf Figure 6}.
N3 is a slight elaboration of N2, and from the standpoint of irrotational area mappings produces the same information
dimension. Here is the {\tt FORTRAN} description of a single step in the corresponding Random Walk :
\begin{verbatim}                                                                                                                  
call random_number(r)                                                                                                             
x = r                                                                                                                             
call random_number(r)                                                                                                             
if (r.lt.(2/3))                   ynew = (y/3) + (0/3)
if((r.ge.(2/3)).and.(r.le.(5/6))) ynew = (y/3) + (1/3)
if (r.gt.(5/6))                   ynew = (y/3) + (2/3)
y = ynew
[ Three-Panel Nonequilibrium Random Walk, N3 ]
\end{verbatim}
Throughout our numerical work we have used the handy FORTRAN random-number generator indicated here, ``{\tt
random}$\_${\tt number(r)}'', said to have a repeat length of order $10^{77}$ (!)

To illustrate the differences in ordering of the bin populations resulting from the first four steps of the random walks
equivalent to N2 and N3 we compare million-point 81-bin histograms of the two Walks in {\bf Figures 7 and 8}.  Evidently the
N3 mapping, starting with a uniform distribution, produces exactly the same set of bin probabilities as does N2 (though
in a different order) and so the walks have exactly the same information dimensions\cite{b24}, 0.789690.  But in the N3 case
the Lyapunov exponents, and the Kaplan-Yorke dimension, are different, and produce an interesting surprise :
$$
\lambda_1 = (2/3)\ln(3/2) + (1/3)\ln(6) =   0.867563 \ ; \
\lambda_2 = (2/3)\ln(1/3) + (1/3)\ln(1/3) =  -1.098612 \ .
$$
$$
\to D_{KY} = 1 - (\lambda_1/\lambda_2) =  1.789690 \ . \ [ \ {\rm Equivalent \ to \ }D_I{\rm \ for \ N3} \ ] \ .
$$
In fact {\it this time} the Kaplan-Yorke conjecture is true, provided one imagines that the steady state maintains the
stationary value of the information dimension observed during the evolution suggested by {\bf Figure 9} !  A proof or
disproof of this thought would be welcome. 

\section{Discussion and Conclusions}
The Kaplan-Yorke conjecture $D_I \stackrel{?}{=}D_{KY}$ is forty years old.  It is surprising that the apparent counterexample
for the linear N2 mapping considered here evaded detection for so many years.  We have seen that the generalized Baker Maps
N2 and N3 agree with both the thermodynamic and the computational statements of the Second Law of Thermodynamics.  The Baker Map
fractals provide exactly the same computational mechanism for dissipation as is present in manybody simulations.  But our
understanding remains incomplete. Why {\it is} or {\it is not} the Kaplan-Yorke approximation valid or invalid for linear maps?
The puzzling difference between pointwise dimensionality and regionwise dimensionality is likewise unsettling, but is firmly
established by our results.

The mathematics of fractal sets remains paradoxical and challenging.  Besides the disagreement between the various versions of
the information dimension the simple geometry of fractals is itself puzzling.  The popular understanding of cumulus clouds as
2.5-dimensional objects suggests that fractals are isotropic.  The Sierpinski carpet and sponge fractals have characteristic
rotational symmetries.  On the other hand {\it all} of the fractals arising from statistical mechanics appear to be anisotropic.
Particular local directions correspond to exponential instability or stability.  Without anisotropicity there could be no
Kaplan-Yorke analysis of dimension.  The unsettling cardinality equivalence of the holy Cantor dust and the holeless continuum,
despite their different information dimensions, suggests that there is much more work for the mathematicians to do, perhaps with
useful physical applications.

In any event a thorough pedagogical explanation of the situation described here will help to understand these issues.  Such an
explanation has been set as the 2020 Ian Snook Prize Problem recently described in Reference 18.

\section{Postscript of 25 April 2020}

We wish to thank the anonymous referee for his many thoughtful suggestions, most of which we have adopted. The
abstract has been completely rewritten. Some repetitions and ambiguities have been removed. Our descriptions of
the Cantor Set and the geometry of time-reversible maps have both been improved. We decided not to follow the
referee's suggestion to add additional references citing work of the many that have been attracted to these and
similar problems independently of our own efforts.  We believe that any attempt at completeness on our part
would have the unintentional effect of slighting others of our kind friends and colleagues.  We have chosen to
refer here only to those works that have directly influenced our own. Comparing the present version to the arXiv
second version of 9 January 2020 will satisfy the curious reader as to all the changes we have made.

\pagebreak

\begin{figure}
\includegraphics[width=2.5 in,angle=-0.]{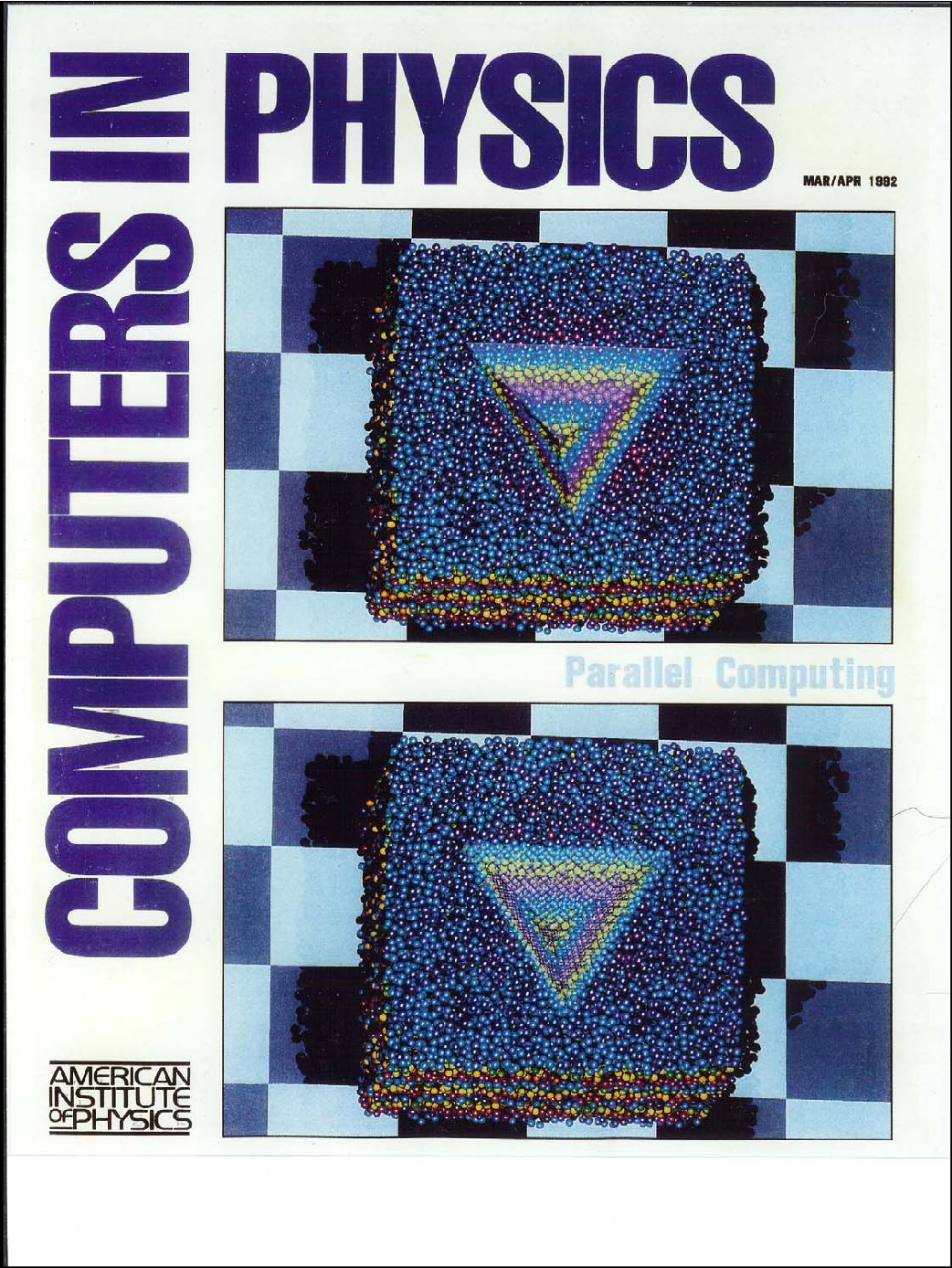}
\caption{
This 1992 cover illustrates two 32 768-atom indentations of amorphous silicon, one with a smooth-faced tetrahedral indentor,
the other with a rough face-centered-cubic atomistic indentor. See Reference 4 for the details.  By 2020 simulations
with trillions of atoms became feasible.
}
\end{figure}

\begin{figure}
\includegraphics[width=2.7in,angle=-0.]{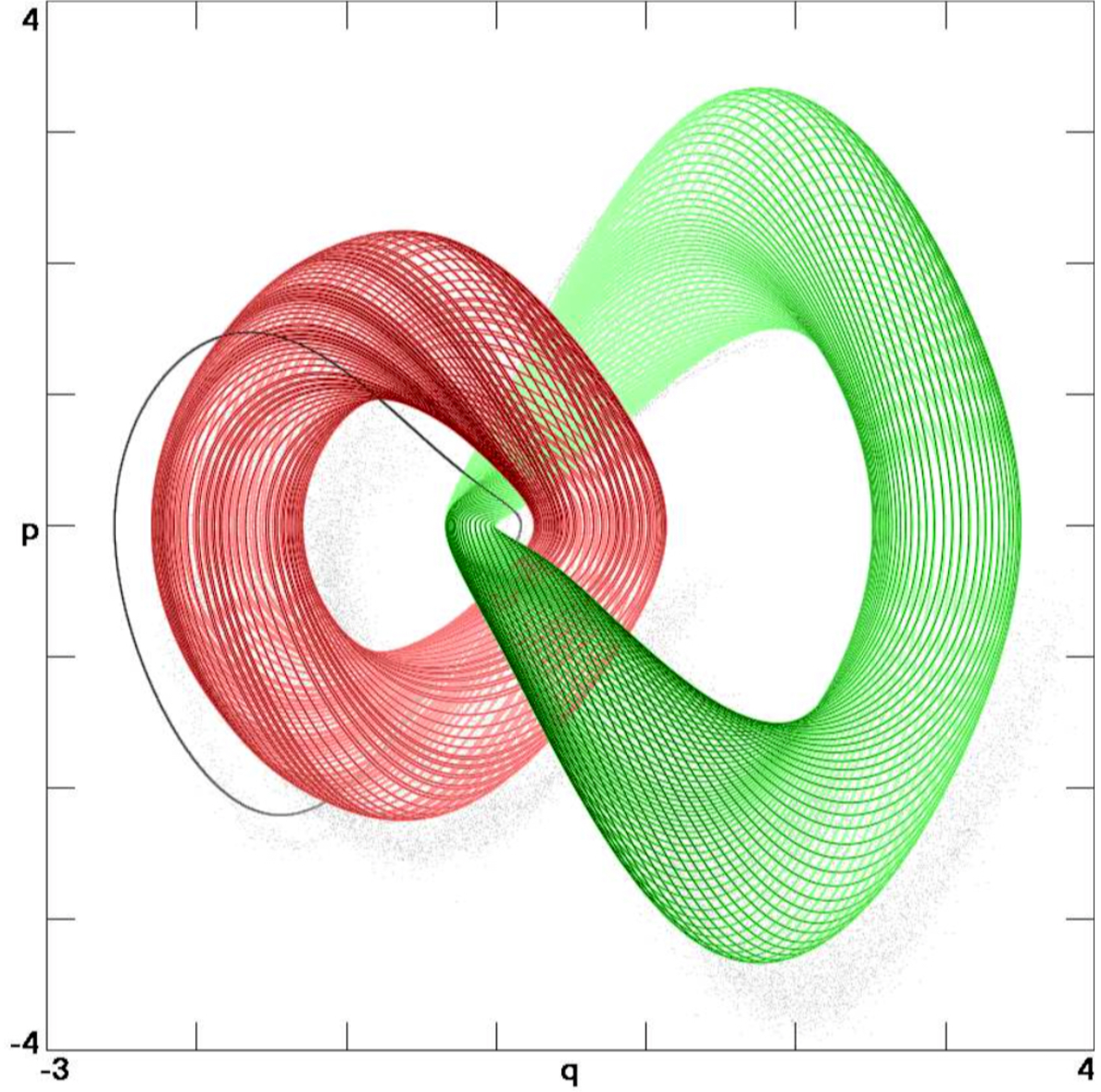}
\caption{
Three stationary solutions for the Nos\'e-Hoover oscillator with maximum temperature gradient $\epsilon = 0.42$.
Unlike the Borromean rings each of the three solutions here is linked to the other two. See Reference 15 for more
details.  The two tori are produced using the initial conditions $(q,p,\zeta)$ = (-2.3,0,0) and (3.5,0,0). The
limit cycle is produced using the initial condition (-2.7,0,0).
}
\end{figure}

\begin{figure}
\includegraphics[width=2.7in,angle=-90.]{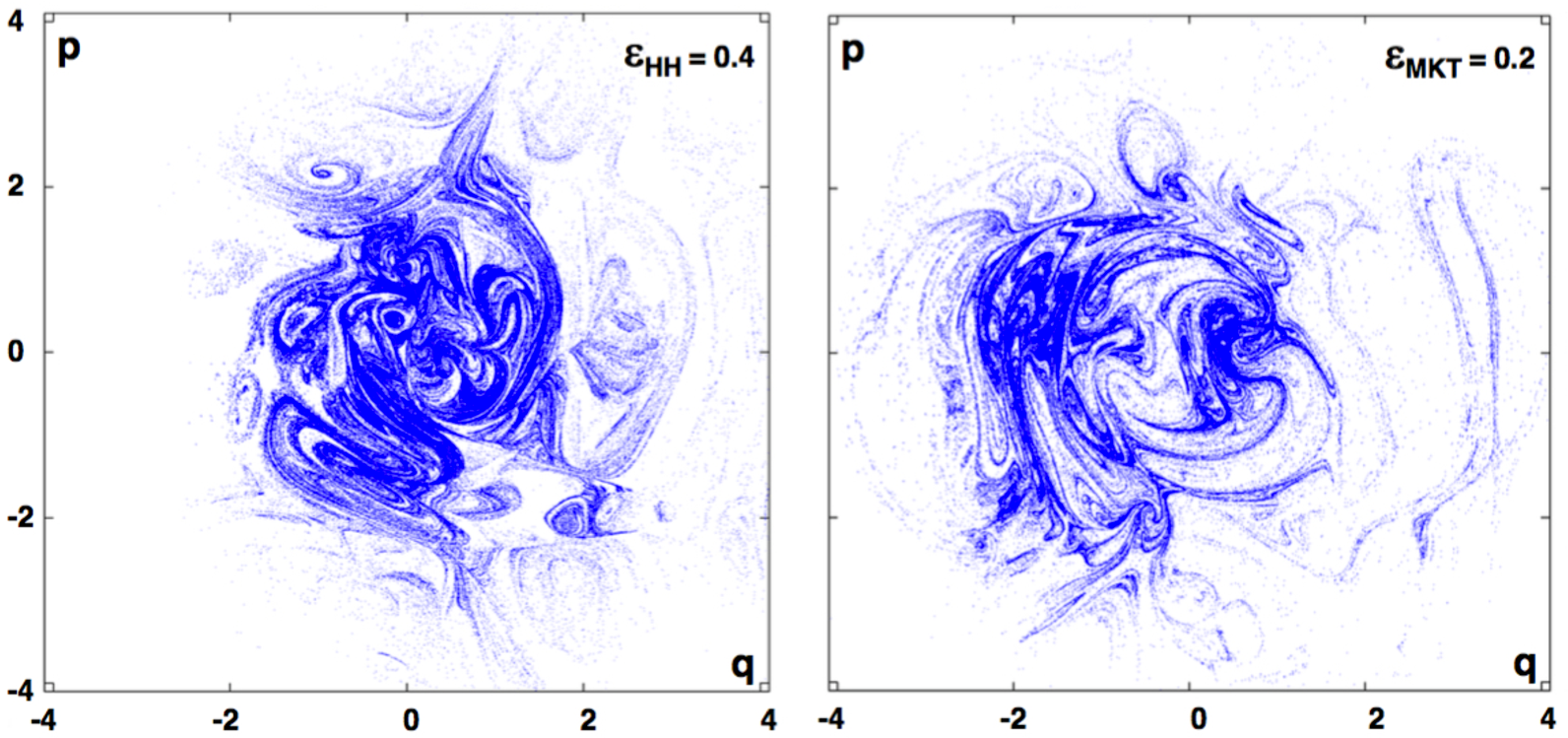}
\caption{
Two fractal $p(q)$ sections near the $\zeta = \xi = 0$ plane for [ left ] the Hoover-Holian Oscillator with
$\epsilon = 0.40$ and for [ right ] the Martyna-Klein-Tuckerman Oscillator with $\epsilon = 0.20$ .
See Section 5.7.2 of W. G. Hoover and C. G. Hoover's {\it Simulation and Control of Chaotic Nonequilibrium
Systems} (World Scientific, Singapore, 2015).  The initial conditions are $(q,p,\zeta,\xi) = (1,0,0,0)$
with every fourth-order Runge-Kutta point plotted satisfying $\zeta^2 + \xi^2 < 0.00001$. Both sections were
generated  with $2\times 10^{11}$ timesteps using $dt= 0.003$.
}
\end{figure}

\begin{figure}
\includegraphics[width=2.5 in,angle=-90.]{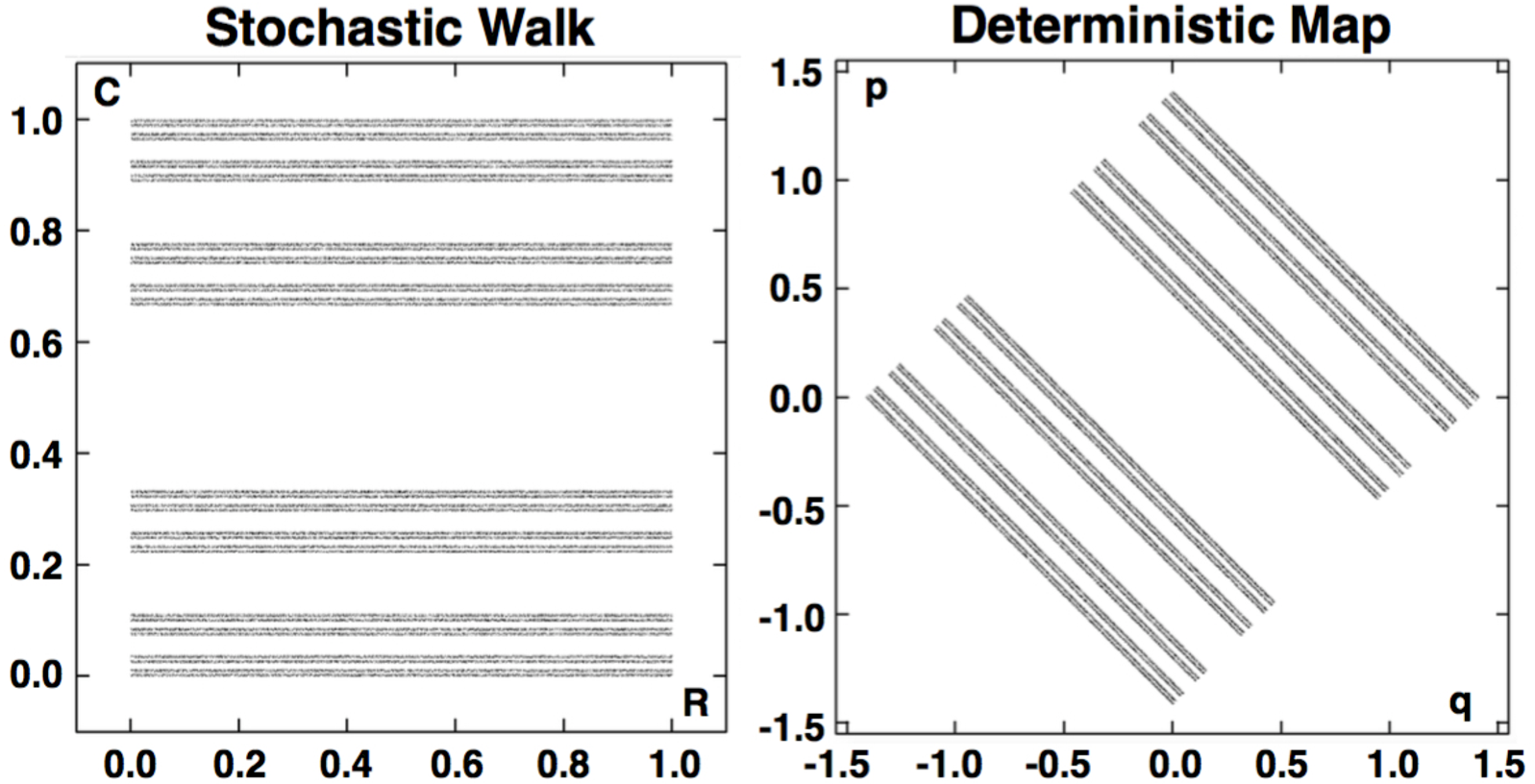}
\caption{
Two two-dimensional forms of the one-dimensional Cantor Set are shown here.  At the left, in the unit square
with $0<R,C<1$ $R$ is chosen randomly and the $C$ coordinate follows a stochastic random walk governed by $R$.
At the right, in the rotated $2\times2$ diamond-shaped domain the next $(q,p)$ point follows from the last
according to the ``Diamond-Shaped Generation'' algorithm given at the end of Section II. The Figure shows
a sequence of 100,000 points in both cases.
}
\end{figure}

\begin{figure}
\includegraphics[width=1.95 in,angle=-90.]{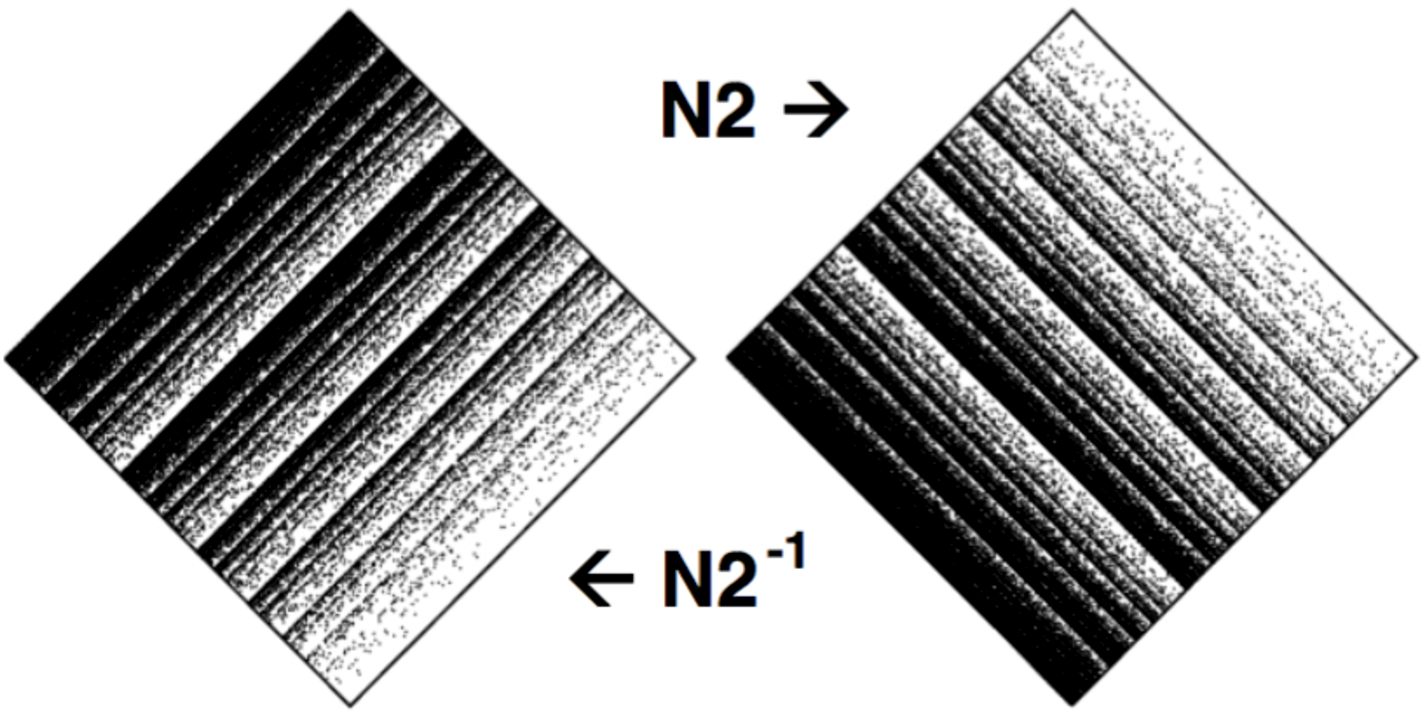}
\caption{
The repellor generated by N2$^{-1}$ (at left) and attractor generated by N2 (at right) using 100 000 iterations
from the initial point $(q,p) = (0,0)$ .  Note that $-\sqrt{2} < q,p <+\sqrt{2}$ . The map is time-reversible so
that the repellor is the mirror-image (with the mirror horizontal) of the attractor.  Although the fractal
dimensions of the attractor and repellor are identical their stabilities (as given by their Lyapunov exponents
from N2) are opposite as a consequence of their time-reversibility.
}
\end{figure}

\begin{figure}
\includegraphics[width=1.5in,angle=-90.]{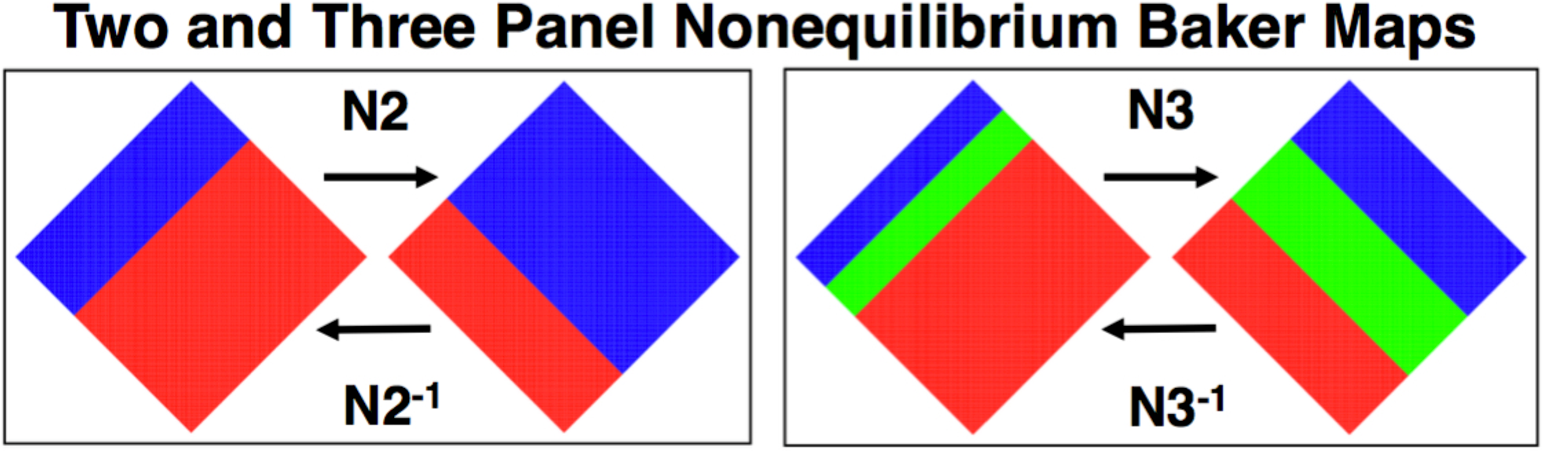}
\caption{
The rotationless two-panel and three-panel maps N2 and N3 are illustrated here.  For more details see our recent
arXiv contributions. N2 is time-reversible with its N2$^{-1}$ = T*N2*T, where the Time-Reversal Mapping T changes
the sign of the vertical ``momentum'', T$(q,\pm p) = (q,\mp p)$.  Although the two mappings are similar N3 is not
time-reversible.
}
\end{figure}

\begin{figure}
\includegraphics[width=4.25in,angle=-90.]{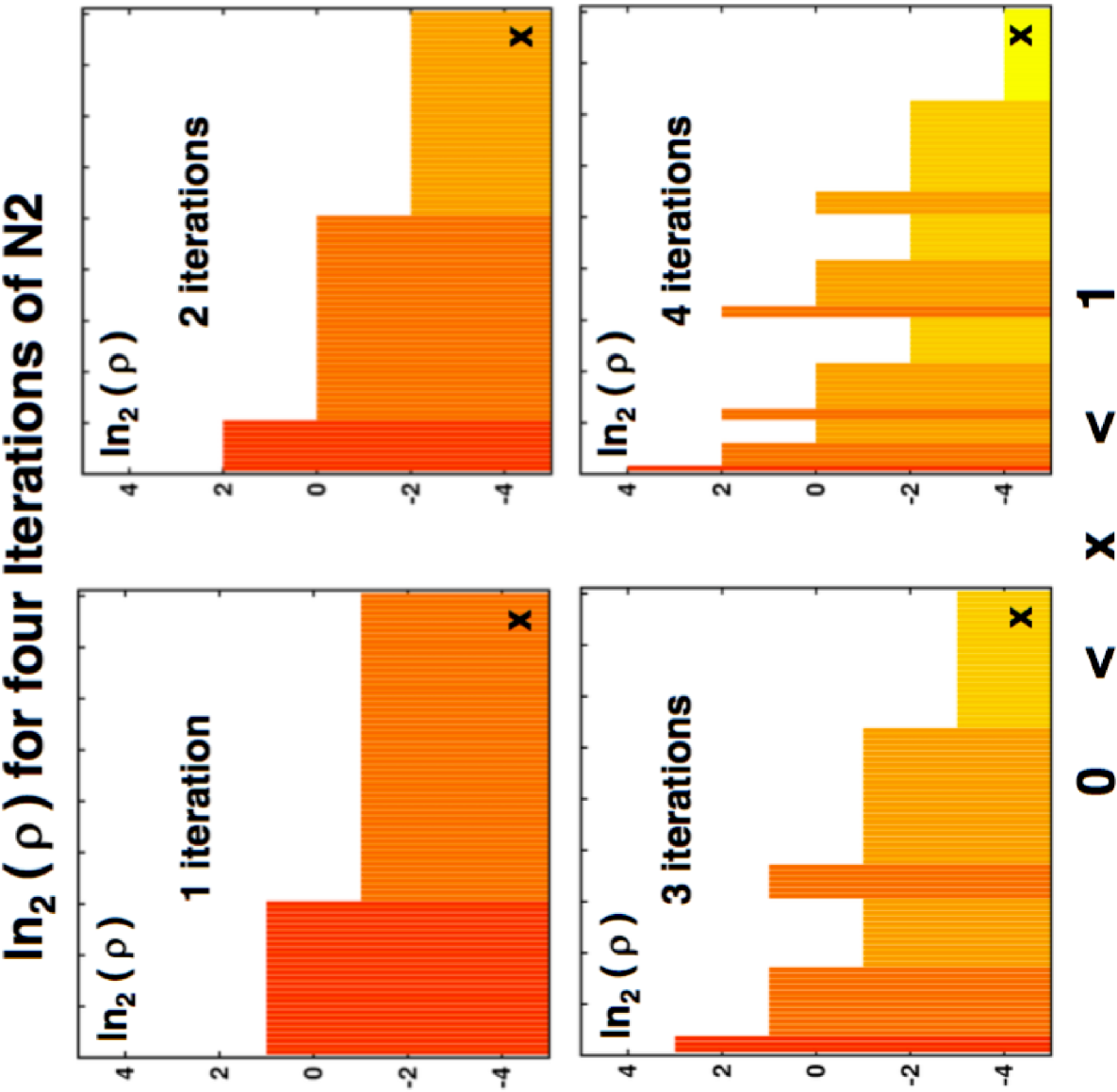}
\caption{
The histograms resulting from four iterations of the random walk version of the N2 map in the unit square. The initial
distribution of one million points was uniform in the unit interval $(0<x<1)$.  After the first iteration, with
one third of the points moving left and two thirds right the probability density is 2 for $(0<x<1/3)$ and (1/2) for
$(1/3<x<2/3)$. Two iterations give probability densities of 4 for $(0<x<1/9)$, 1 for $(1/9<x<5/9)$ and (1/4) for
$(5/9<x<1)$. After the fourth iteration the numbers of bins at each level of probability density, from the highest, 16,
to the lowest, 1/16, are $\{ \ 1\times 1, 2\times 4, 4\times 6, 8\times 4, 16\times 1 \ \}$, 81 in all.
}
\end{figure}

\begin{figure}
\includegraphics[width=4.25in,angle=0.]{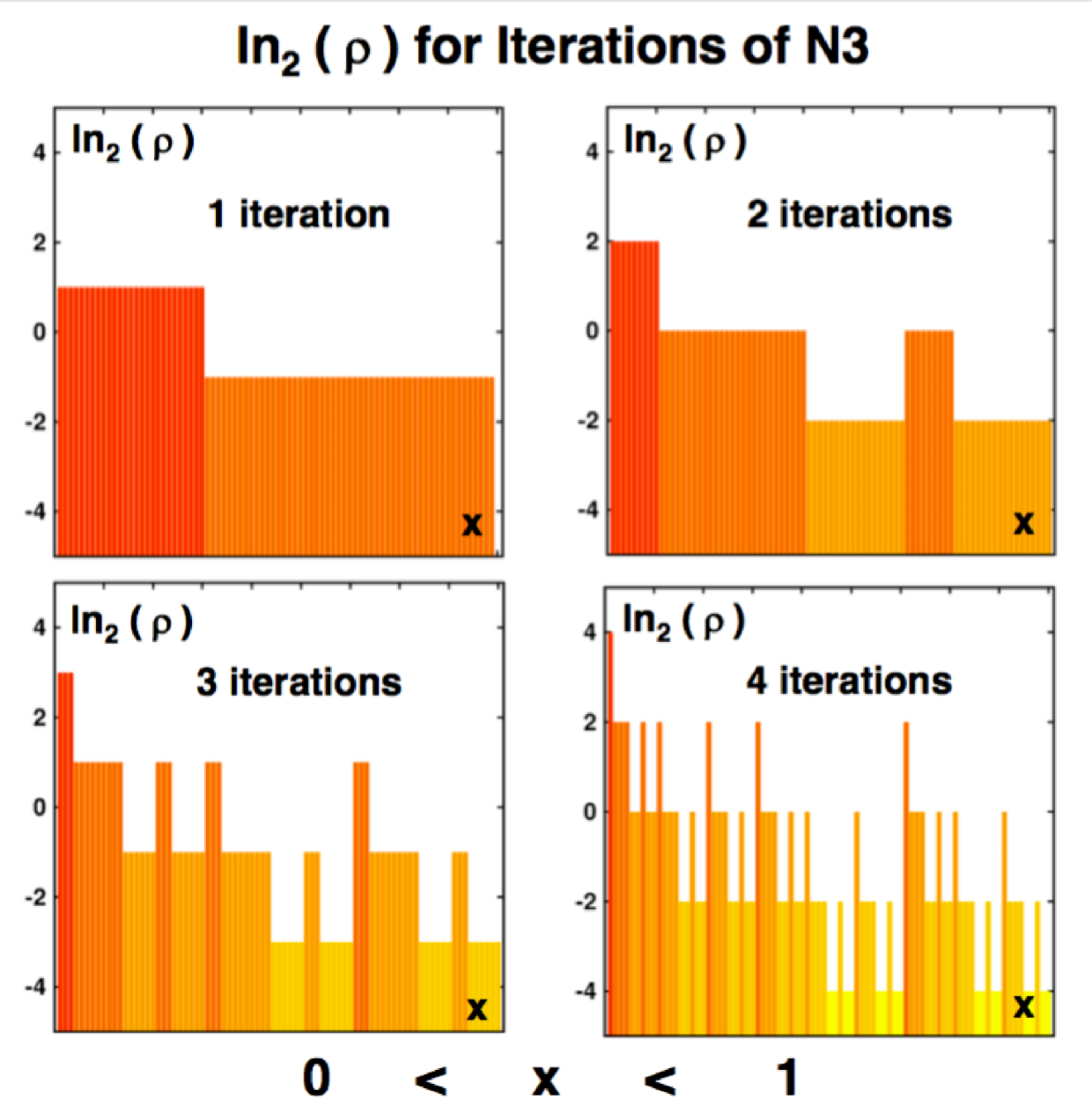}
\caption{
The histograms resulting from four iterations of the random walk version of the N3 map in the unit square. The initial
distribution of one million points was uniform in the horizontal interval $(0<x<1)$. The histograms show the base-2
logarithms of the fraction of the points in each of 81 bins of equal width. These bin probabilities are equal here to
probability densities which integrate to unity over the interval $(0<x<1)$. We show five different probability-density
levels from $2^4$ for the highest probability density to $2^{-4}$ for the least probable. After the fourth iteration
the numbers of bins at the five different levels, from the highest probability to the lowest, are exactly the same as
those from the N2 mapping in {\bf Figure 7}. 
}
\end{figure}

\begin{figure}
\includegraphics[width=2.7in,angle=-90.]{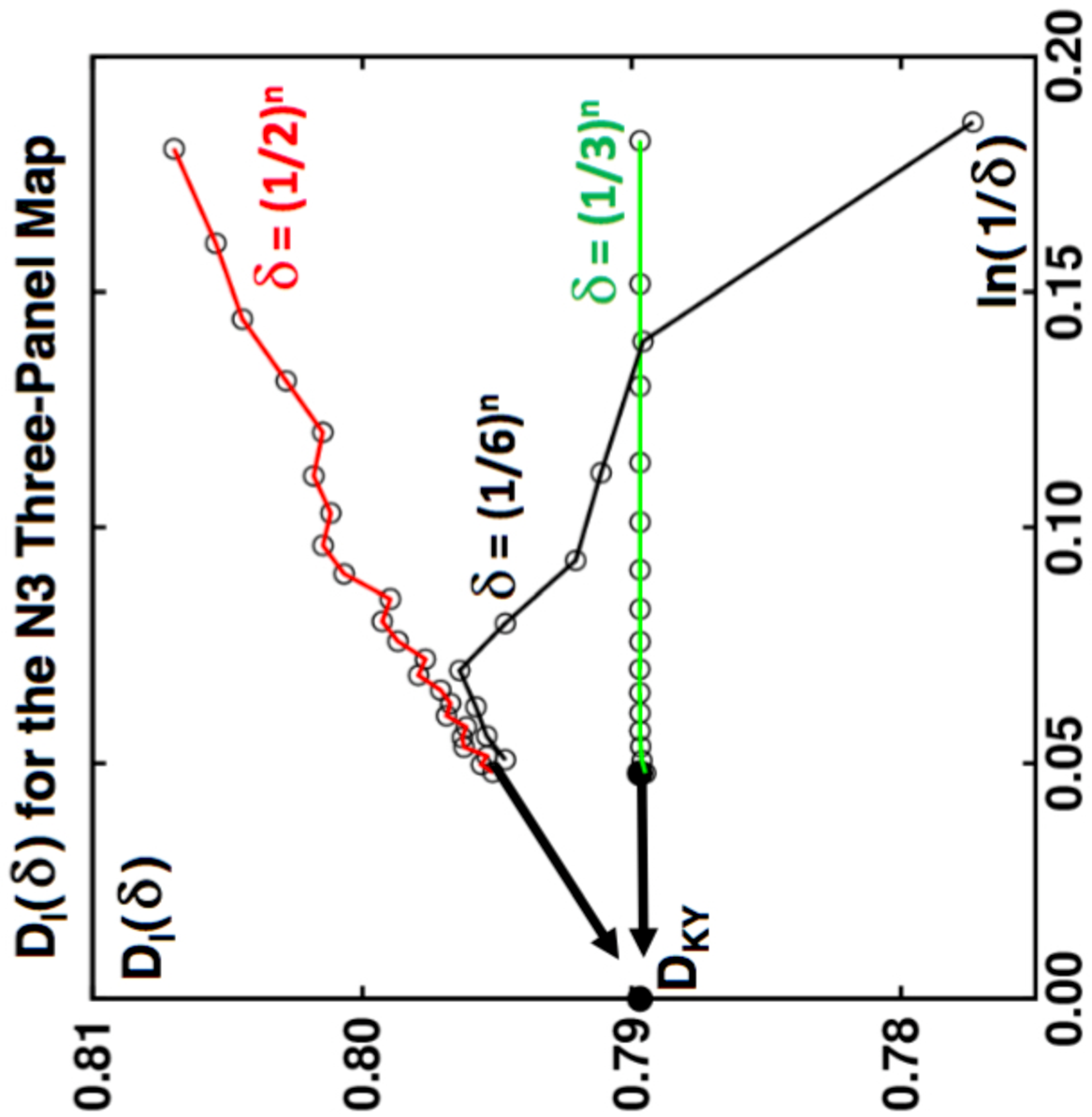}
\caption{
Information Dimension data for the random walk problem equivalent to the N3 Mapping are all consistent with the same
information dimension $D_I = D_{KY} = 0.789690$ for the Walk and 1.789690 for the Mapping. Pointwise analyses of the
N3 mapping are described here for meshes of $(1/2)^n$ (red), $(1/3)^n$ (green), and $(1/6)^n$ (black).  Analogous data
for the N2 mapping are displayed in Figure 5 of Reference 18.
}
\end{figure}

\end{document}